\documentclass{emulateapj}
\usepackage{apjfonts}
\usepackage{graphicx}
\usepackage{epsf}
\usepackage{epsfig, natbib, graphicx, color}
\usepackage{hyperref}

\newcommand{\avg}[1]{\langle#1\rangle}
\newcommand{\Mobs}{M_{\rm obs}}

\newcommand{\hiMsun}{h^{-1}M_\odot}

\begin{document}
\journalinfo{The Astrophysical Journal, 713:856--864,  2010 April 20}
\submitted{Received 2009 November 4; accepted  2010 March 1; published 2010 March 26}
\shortauthors{WU, ZENTNER \& WECHSLER}
\shorttitle{REQUIRED ACCURACY OF THE HALO MASS FUNCTION AND BIAS}

\title{The Impact of Theoretical Uncertainties in the Halo Mass
Function and Halo Bias on Precision Cosmology}

\author{Hao-Yi Wu\altaffilmark{1}, Andrew R. Zentner\altaffilmark{2},
Risa H.  Wechsler\altaffilmark{1}}
\altaffiltext{1}{Kavli Institute for Particle Astrophysics and
Cosmology, Physics Department, and SLAC National Accelerator
Laboratory, Stanford University, Stanford, CA 94305, USA}
\altaffiltext{2}{Department of Physics and Astronomy, University of
Pittsburgh, Pittsburgh, PA 15260, USA}

\begin{abstract}

We study the impact of theoretical uncertainty in the dark matter halo
mass function and halo bias on dark energy constraints from imminent
galaxy cluster surveys.  We find that for an optical cluster survey
like the Dark Energy Survey, the accuracy required on the predicted
halo mass function to make it an insignificant source of error on dark
energy parameters is $\approx 1\%$.  The analogous requirement on the
predicted halo bias is less stringent ($\approx 5\%$), particularly if
the observable--mass distribution can be well constrained by other
means. These requirements depend upon survey area but are relatively
insensitive to survey depth.  The most stringent requirements are
likely to come from a survey over a significant fraction of the sky
that aims to observe clusters down to relatively low mass, $M_{\rm th}
\approx 10^{13.7}\ \hiMsun$; for such a survey, the mass function and
halo bias must be predicted to accuracies of $\approx 0.5\%$ and
$\approx 1\%$, respectively.  These accuracies represent a limit on
the practical need to calibrate ever more accurate halo mass and bias
functions.  We find that improving predictions for the mass function
in the low-redshift and low-mass regimes is the most effective way to
improve dark energy constraints.

\end{abstract}

\keywords{cosmological parameters --- cosmology: theory --- galaxies: clusters --- galaxies: halos}

\section{Introduction}

Observations of the number density of galaxy clusters as a function of
cluster mass and redshift are powerful probes of cosmology, especially
the accelerated cosmological expansion caused by the cryptically
dubbed {\em dark energy}. Cosmological parameters have been recently
inferred from a number of observations employing different techniques
for cluster identification
\citep[e.g.,][]{Gladders07,Mantz08,Henry09,Rozo09,Vikhlinin09,Mantz09a}. The
effort to derive precise and unbiased constraints on cosmological
parameters, particularly those describing dark energy, requires
control of various systematic error sources.  In this paper, we
estimate the precision with which the abundance of dark matter halos
as a function of mass (the halo {\em mass function}) and the
clustering of halos as a function of mass (the halo {\em bias}) must
be predicted in order to ensure that errors in the predictions for
these quantities will be a small fraction of the error budget in
forthcoming cluster count cosmology efforts.

For the current generation of cluster count surveys, such as the Sloan
Digital Sky Survey (SDSS), the Red-Sequence Cluster Survey (RCS), and
the Massive Cluster Survey (MACS), uncertainty in the mass function is
unlikely a major concern.  The dominant errors in contemporary surveys
are thought to be either limited statistics \citep[e.g.,][for
MACS]{Mantz08} or the uncertain relation between observable quantities
and cluster mass \citep[e.g.,][for RCS and SDSS
respectively]{Gladders07,Rozo09}.  Uncertainty in the relation between
cluster observables and masses will be an important limitation
\citep[e.g.,][]{Majumdar03,Majumdar04,LimaHu04,LimaHu05,Stanek06,LimaHu07,Cunha08},
and it is thought that some combination of empirical and theoretical
insight into the observable--mass relation will continue to prove
effective in the interpretation of forthcoming data
\citep[e.g.,][]{Kravtsov06, Rozo09Richness}.  Understanding the
relation between observables and mass is a rapidly evolving subject,
so it is difficult to anticipate the level at which this will be
controlled in the analyses of forthcoming data.  Although we do not
directly deal with observable--mass relations, we explore the
dependence of our results on assumptions about the observable--mass
relation.

The motivations to study the uncertainty in predicting halo abundances
and clustering are twofold.  First, forthcoming optical surveys, such
as the Dark Energy Survey (DES)\footnote{ {\tt
http://www.darkenergsurvey.org}}, the Panoramic Survey Telescope \&
Rapid Response System (PanSTARRS)\footnote{ {\tt
http://pan-starrs.ifa.hawaii.edu}}, and the Large Synoptic Survey
Telescope (LSST)\footnote{ {\tt http://www.lsst.org}}; X-ray surveys,
such as the extended ROentgen Survey with an Imaging Telescope Array
(eRosita)\footnote{ {\tt
http://www.mpe.mpg.de/projects.html\#erosita}}; and Sunyaev--Zeldovich
(SZ) surveys, such as the Atacama Cosmology Telescope (ACT)\footnote{
{\tt http://www.physics.princeton.edu/act/}} and the South Pole
Telescope (SPT)\footnote{ {\tt http://pole.uchicago.edu/}}, all
promise to expand greatly upon contemporary observations of
clusters. These surveys will enable significant reductions in
statistical errors and provide better constraints on the cluster
observable--mass distribution.  With ever-improving theoretical
understanding of this distribution, systematic errors that are
currently unimportant may be damaging to future surveys and need to be
controlled.  For example, \citet{Crocce09} demonstrated that current
errors in predicted mass functions may lead to statistically
significant systematic errors in the inferred dark energy equation of
state from SZ surveys.

Second, there is a significant, ongoing effort to develop ever more
accurate predictions for the halo mass function
\citep[e.g.,][]{ShethTormen99,Sheth01,Jenkins01,Evrard02,Reed03,
Warren06,Lukic07,CohnWhite08,Tinker08,Robertson09,Lukic09,Crocce09}
and halo bias
\citep[e.g.,][]{ShethTormen99,Sheth01,SeljakWarren04,Tinker08,Tinker09}. Mass
functions and halo biases have been derived on theoretical grounds
\citep[][see \citealt{Zentner07} for a recent review]
{PressSchechter74,Bond91,MoWhite96,ShethTormen99,Sheth01,
Maggiore09a,Maggiore09b,Maggiore09c}.  Fits to numerical simulations
with theoretically motivated functional forms have been provided by
several groups, and the current state-of-the-art includes the recent
papers by \citet{Lukic07}, \citet{Tinker08}, and \citet{Crocce09}, as
well as the forthcoming work of the Los Alamos National Laboratory
group (S. Bhattacharya et al. 2010, in preparation).  Current
theoretical predictions of the mass function are accurate at the $\sim
10\%-30\%$ level
\citep[e.g.,][]{Tinker08,Robertson09,Rudd08,Stanek08}, and future
efforts will continue to improve this accuracy. These efforts require
numerous cosmological numerical simulations along with requisite
verification, validation, and considerable analysis, involving large
commitments of both computational and human resources.  Therefore,
establishing the accuracy in the mass function and halo bias that are
required by surveys is a particularly timely issue.

In this work, we quantify how accurately halo mass functions and
biases need to be predicted in order to have a negligible contribution
to the error budgets of forthcoming cluster surveys. We begin in
Section~\ref{sec:error} with a simple demonstration of the potential
errors that may be induced by the inaccuracy in the mass function.  We
present our mass function parameterizations and cosmological parameter
constraint forecasts in Section~\ref{sec:parameterization} and
Section~\ref{sec:implementation}, respectively.  In all cases, we take
the \citet{Tinker08} mass function and \citet{Tinker09} halo bias as a
fiducial model, about which we perturb to estimate errors relevant to
forthcoming optical and SZ surveys.

We detail our results in Section~\ref{sec:results}. In summary, we
find that for relatively near-term optical surveys like DES, or SZ
surveys like SPT, the mass function must be calibrated to the $\approx
1\%$ level while the halo bias must be calibrated to the $\approx 5\%$
level.  The requirement on the mass function is relatively insensitive
to different assumptions about the observable--mass distributions
within the range considered in contemporary work, whereas the
requirement on the halo bias is more sensitive.  For longer-term
experiments, such as a nearly half-sky optical survey planned for
LSST, the calibration must be improved to the $\approx 0.5\%$ level in
the mass function and the $\approx 1\%$ level in the halo bias.  This
represents the most stringent requirement for the surveys being
planned over the next decade and may serve as a practical endgame in
the need to refine theoretical predictions for halo abundances and
clustering.

We also explore the halo masses and redshifts at which it is most
important to make accurate predictions for the halo mass function.  We
find that the most effective strategy to improve dark energy
constraints is to improve predictions at the low masses and low
redshifts involved in the survey.  Other details of the dependence of
calibration requirements on mass and redshift depend upon the
observable--mass distribution.  We summarize our results and draw our
conclusions in Section~\ref{sec:conclusion}.

As we were completing our study, we learned of the related work of
\citet{Cunha09MF}. Although our results generally agree with the
results in this study, our study has a number of distinct and
complementary aspects. These authors concentrated on the specific
parameters in the \citet{Tinker08} mass function and the
\citet{ShethTormen99} halo bias, focusing considerable study on the
degeneracies between nuisance parameters and cosmological parameters.
In contrast, we aim to describe the uncertainty in mass function and
halo bias in a manner that is independent of the form of any specific
fitting function. We discretize the mass function and halo bias, and
assign to each mass and redshift bin a distinct nuisance parameter
specifying the fractional deviation of the mass function or halo bias
from the fiducial values. This parameterization also allows us to
identify the masses and redshifts at which it is most fruitful to
refine predictions in order to limit systematic errors on dark energy
parameters.

\section{The Effect of Inaccuracy in the Halo Mass Function}
\label{sec:error}

\begin{figure}[t!]
\epsscale{1.2}
\plotone{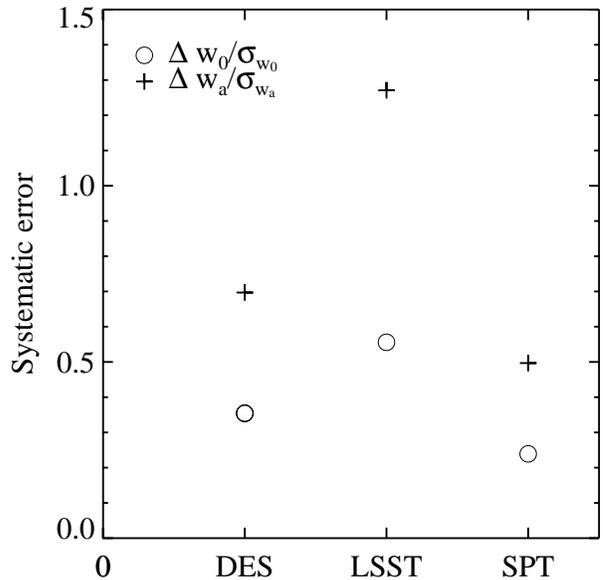}
\caption{Potential systematic errors in dark energy parameter
estimates caused by inaccurate modeling of the mass function. The
systematic errors are shown in units of the statistical uncertainties
expected from each survey. We assume that the fitting formula in
\cite{Tinker08} describes the data, while that in \cite{ShethTormen99}
is used in the likelihood function. The effect on imminent cluster
surveys is not negligible, especially for a wide-area survey like
LSST. Moreover, this systematic error is representative of current
limitations of $N$-body methods, while baryonic processes are known to
cause shifts in halo mass functions of equal or greater magnitude
\citep{Rudd08,Stanek08}.}
\label{fig:Surveys}
\end{figure}

We begin with an example showing the influence of inaccurate modeling
of the halo mass function on the cosmological parameters derived from
galaxy cluster surveys. We compare two commonly used fitting functions
derived from $N$-body simulations, the \citet{ShethTormen99} and
\citet{Tinker08} mass functions. To construct a concrete example, we
assume that the \citet{Tinker08} function describes the true halo mass
function, while the \citet{ShethTormen99} function serves as an
example of an imperfect model in our likelihood analysis. The
differences between these functions are $\sim 20\%$ over the most
relevant masses \citep[see][]{Tinker08}. This represents a reasonable
assessment of our current ability to predict the mass function. For
example, \citet{Tinker08} and \citet{Robertson09} discussed
fundamental limitations in the ability of contemporary methods to
construct analytical mass function fits at the $\sim 10\%$ level, and
baryonic processes can give rise to additional deviations in predicted
mass functions as large as $\sim 30\%$ \citep{Rudd08,Stanek08}.

We use the Fisher matrix formalism for parameter constraint
forecasting to compute the systematic error in dark energy parameters
due to a mass function that is imperfectly calibrated.  We use the
term ``systematic error'' to describe the offset between inferred
values of cosmological parameters and the true values of these
parameters.  This systematic error is often referred to as a ``bias''
in the literature, but we do not use this term in order to avoid
confusion with the halo bias.  We refer the reader to the appendix of
\cite{WuHY08} for the derivation of the systematic errors relevant for
cluster surveys and defer further discussion of our methods to the
following sections.  We describe dark energy by a two-parameter
equation of state $w(a)=w_0 + (1-a)w_a$ and take $w_0$ and $w_a$ as
free parameters.

Figure~\ref{fig:Surveys} shows the systematic errors in dark energy
parameters in terms of the statistical uncertainties.  We present the
absolute value of the error, and zero indicates that there is no
systematic error.  Three survey assumptions are shown: a DES-like
optical survey ($5000 \deg^2$ to a maximum redshift $z_{\rm max}=1$),
an LSST-like optical survey ($20,000 \deg^2$ to $z_{\rm max}=1$), and
an SPT-like SZ survey ($2000 \deg^2$ to $z_{\rm max}=1.5$).  As can be
seen, the dark energy equation of state parameters may have systematic
errors at levels of approximately 20\%--130\% of their statistical
uncertainties.  These errors indicate that current inaccuracy in the
halo mass function would not be a negligible contribution to the
errors in dark energy parameters inferred from the next generation of
surveys.  In fact, this is likely a conservative estimate of the
systematic error because it does not include additional uncertainties
induced by baryonic physics.

\section{Methods}

\subsection{Parameterizing Uncertainty in the Halo Mass Function and Halo Bias}
\label{sec:parameterization}

In this subsection, we describe our parameterization of an uncertain
halo mass function $d n/d \ln M$ and an uncertain halo bias $b(M)$,
aiming to decouple our results from the particulars of published fits
to halo mass functions and biases.  For the mass function, we define a
set of nuisance parameters $f_i$ as follows. We take the ratio of the
actual mass function to a fiducial mass function at mass $M$ and
redshift $z$ to be
\begin{equation}\label{eq:fi}
\frac{dn/d\ln M}{(dn/d\ln M)_{\rm Fid}} = f_i \psi_i(M,z) \ ,
\end{equation}
where the $\psi_i(M,z)$ describes the binning in mass and redshift and
equals unity when $(M,z)$ is within the range specified for bin $i$.
The index $i$ runs over all bins in both the mass and redshift
dimensions.

As we will discuss in more detail in Section~\ref{sec:implementation},
galaxy cluster surveys observe halo mass proxies above some observable
threshold. In what follows, we model the theoretical mass function
down to a minimum mass that is lower than the observable threshold by
an amount comparable to the scatter in the observable--mass
distribution ($\sigma_{\ln M}$, which will be defined later). This low
minimum mass is necessary to account for the scatter, which may lead
relatively low-mass clusters with higher-than-average mass proxies to
be included in an observed cluster sample.

Between the minimum mass and a maximum mass, we use several bins for
the theoretical mass function.  We choose the width of the bins to be
roughly comparable to the scatter in the relation between observable
and mass.  This scheme follows from two considerations.  First, the
binning in the theoretical mass function should be sufficiently fine
so that constraints on dark energy are maximally degraded when there
is no prior information on any of the $f_i$ nuisance parameters.  In
this way correlated shifts in the mass function over a range of masses
do not significantly aid in self-calibration.  Second, very fine mass
bins are unnecessary because the scatter in the observable--mass
relation gives an effective resolution with which the mass function
may be probed.  Binning significantly more finely means that several
theoretical mass bins contribute to a measurement at a particular
observable mass proxy.  The value of the mass function in each bin
will then need to be predicted with less accuracy, but the combination
that results in the measurement will need to be predicted with fixed
accuracy.

Following from these considerations, it is clear that mass function
accuracy requirements should depend upon bin size.  We quote results
that are realized when the mass function is binned at a resolution
comparable to the scatter in the relation between observable and mass.
This seems sensible in that it allows for maximal degradation of
cosmological constraints when the mass function is uncertain, but does
not sample the mass function significantly more finely than
observations.

For our model optical surveys, this choice amounts to five bins spaced
evenly in $\ln M$ between $M_{\rm min}=10^{13.3}\ \hiMsun$ and $M_{\rm
max}=10^{15.3}\ \hiMsun$.  For our model SZ surveys, we use seven mass
bins above $M_{\rm min}=10^{13.9}\ \hiMsun$. Note that these binning
methods for $\ln M$ include one bin below the threshold of observable
mass proxies discussed in the next section. We bin in redshift
intervals of $\Delta z=0.1$ from $z=0$ to a maximum survey redshift,
$z_{\rm max}$.  For our fiducial mass function, we use the mass
function fit at a spherical overdensity of $\Delta = 200$ in
\citet{Tinker08}.

We introduce an analogous set of parameters $g_i$ to describe the uncertainty in halo
bias around the fiducial value from \cite{Tinker09}, 
\begin{equation}\label{eq:gi}
\frac{b(M,z)}{b _{\rm Fid} (M,z)} = g_i \psi_i(M,z) \ ,
\end{equation}
according to a binning scheme identical to that of the mass function.
The halo bias and mass function may well be linked based on physical
considerations \cite[e.g., through the excursion set
approach]{Kaiser84,MoWhite96,ShethTormen02,Zentner07}, but we treat
them independently as a conservative way to account for the inadequacy
of the assumptions of particular models.

In our statistical forecasts, we include the parameters $f_i$ and
$g_i$ at each of the set mass and redshift bins. We then determine how
well we must be able to set priors on the parameters $f_i$ and $g_i$.
We refer to the prior knowledge of the $f_i$ and $g_i$ as $\sigma_{\rm
f_i}$ and $\sigma_{\rm g_i}$ respectively.

\subsection{Implementation}
\label{sec:implementation}

We forecast constraints on cosmological parameters using the methods
described in detail in \citet{WuHY08,WuHY09}. In what follows, we give
a brief description of the aspects of this calculation necessary to
quantify the effect of uncertainty in the mass function and halo bias
and refer the reader to the aforementioned papers for a more complete
description.

We assume that a survey determines a particular mass proxy $\Mobs$ and
that the survey clusters are binned in $\ln \Mobs$ according to a
binning function $\phi(\ln \Mobs)$. We assume that mass proxies are
related to mass through an observable--mass distribution $P(\ln
\Mobs|\ln M)$ which specifies the probability that a halo with natural
logarithm of mass $\ln M$ gives a mass proxy $\ln \Mobs$. As an
example, consider a survey composed of a number of cells of volume
$V$, each of which spans a narrow range in redshift. The mean number
of clusters in each cell is
\begin{eqnarray}
\bar{m} &=&  V \int d\ln M \left(\frac{d n}{d \ln M}\right) \avg{\phi|\ln M} \nonumber\\
&=& V \sum_{k} \int d \ln M f_k \psi_k(M,z)
\left(\frac{d n}{d \ln M}\right)_{\rm Fid} \avg{\phi|\ln M}
\end{eqnarray}
where
\begin{equation}
\avg{\phi|\ln M} \equiv \int d \ln \Mobs P(\ln \Mobs | \ln M ) \phi(\ln \Mobs).
\end{equation}
and the index $k$ runs over all mass bins in the theoretical mass
function at the corresponding redshift.

We apply counts-in-cells and self-calibration as in \cite{LimaHu04,
LimaHu05}, calculating both the expected counts in different bins,
$\bold{\bar m}$, and the variance in counts, $\bold S = \langle (\bold
m-\bold{\bar m})^T(\bold m-\bold{\bar m}) \rangle$. The Fisher matrix
for a cluster survey is given by
\begin{equation}
F_{\alpha \beta}=\bold{\bar m}^{T}_{,\alpha} \bold C^{-1} \bold{\bar m}_{,\beta} 
+ \frac{1}{2} {\rm Tr}[\bold C^{-1} \bold S_{,\alpha}\bold C^{-1} \bold S_{,\beta}]
\ ,
\end{equation}
where $\bold C = {\rm diag} (\bold{ \bar m}) + \bold S$ is the total
covariance matrix.  The comma followed by a Greek letter subscript
refers to the derivatives with respect to model parameters, which
include multiple $f_i$ and $g_i$ in this analysis.  With this
notation, the derivative of the mean cluster count in a particular
cell with respect to $f_i$ reads
\begin{equation}
\bar{m},_{\rm f_i} = \frac{\partial\bar m}{\partial f_i} = V\int d\ln M \psi_i(M)
\left(\frac{d n}{d \ln M}\right)_{\rm Fid} \avg{\phi|\ln M}.
\end{equation}
Analogous expressions hold for the variance in counts.

We assume an observable--mass distribution $P(\ln \Mobs | \ln M)$ that
is a Gaussian with a mean $\avg{\ln \Mobs|\ln M } = \ln M + \ln M_{\rm
bias}$ and a variance $\sigma_{\ln M}^2$. We parameterize the mean and
variance of our observable--mass distribution as in \cite{LimaHu05},
with an additional mass dependence:
\begin{eqnarray}
\ln M_{\rm bias} &=& \ln M_0+\alpha_M\ln (M/M_{\rm pivot})+\alpha_z \ln (1+z) \label{eq:Mbias}\nonumber \\
\sigma_{\ln M}^2 &=& \sigma_{\rm Fid}^2+ \beta_M\ln (M/M_{\rm pivot})+B_0 +
B_1 z + B_2 z^2+ B_3 z^3\  
\label{eq:scatter}
\end{eqnarray}
Here $M_{\rm pivot}$ is the pivot mass characterizing the mass
dependence.  We fix its value to the observable threshold of the
survey and note that the uncertainty in $M_{\rm pivot}$ can be
accounted for with $B_0$ and $\ln M_0$ so that this choice is
innocuous.  The parameter $\sigma_{\ln M}$ is referred to as the
observable--mass scatter throughout this work.  This model includes
eight nuisance parameters in total: $(\ln M_0,\ \alpha_M ,\ \alpha_z,\
\beta_M,\ B_0,\ B_1, \ B_2,\ B_3)$; their fiducial values are all
assumed to be zero.

\begin{figure*}[t!]
\epsscale{1.2}
\plotone{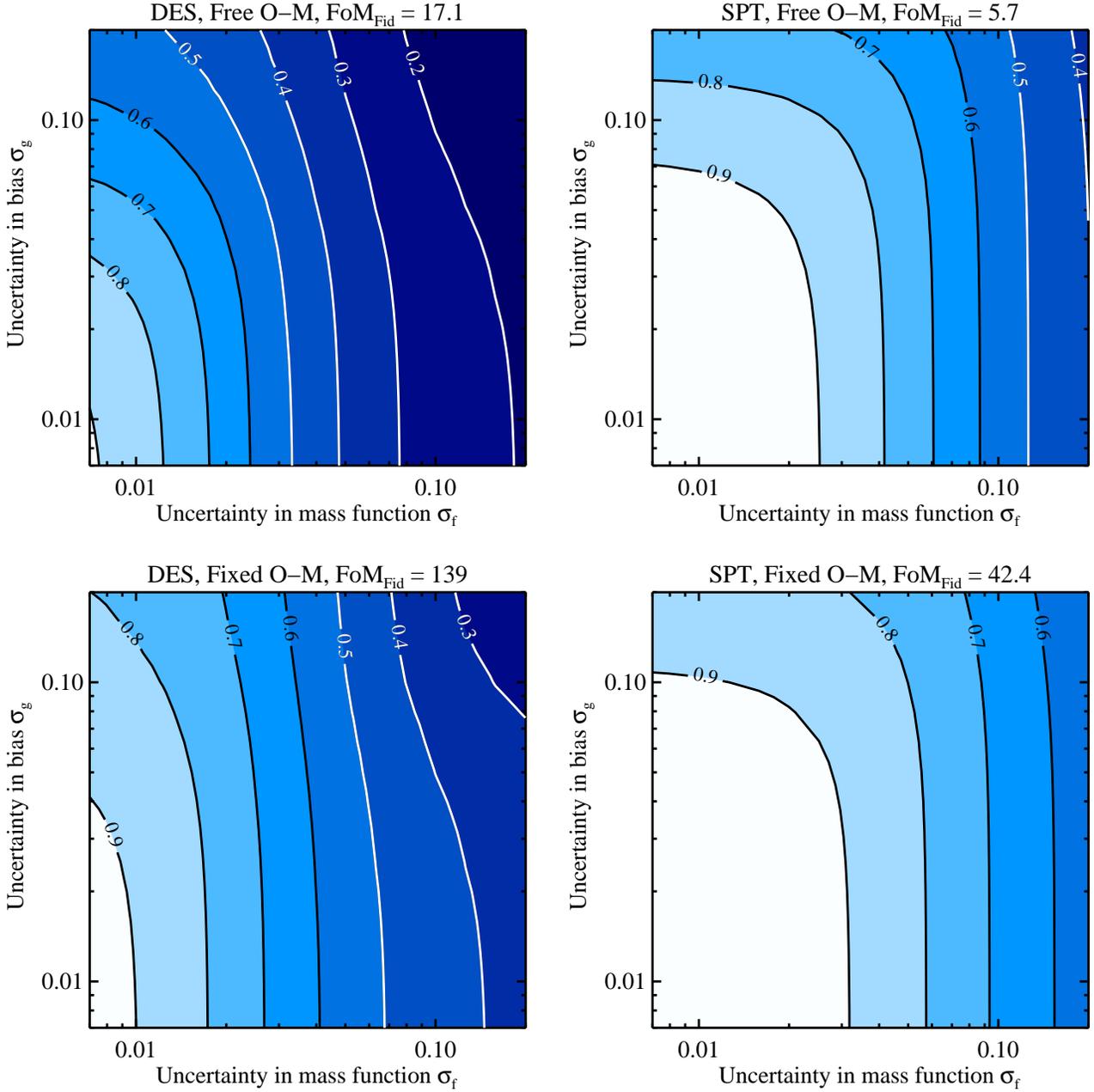}
\caption{ Degradation of the FoM due to uncertainties in mass function
and halo bias.  The contours and numbers correspond to the degraded
FoM with respect to the FoM with perfectly-known mass function and
halo bias.  In order to quote a relatively simple result, we assume
that the mass function parameters all have the same prior,
$\sigma_{\rm f}$, and that the halo bias parameters all have the same
prior, $\sigma_{\rm g}$.  Given the statistical power of DES- and
SPT-like surveys, the mass function needs to be predicted with a few
percent precision to avoid $10\%$ degradation in the FoM.  For a
perfectly known observable--mass distribution, the required precision
of halo bias is less stringent because the information from sample
variance becomes less important.  Comparing a DES-like survey and an
SPT-like survey, the latter has less stringent requirements because
its smaller sky coverage and higher observable threshold result in
fewer observed clusters.}
\label{fig:f_g} 
\end{figure*}

\begin{figure}[t!]
\epsscale{1.2}
\plotone{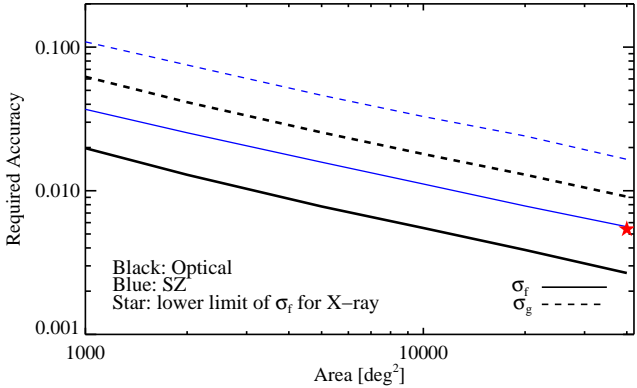}
\caption{ Effect of survey area on the required accuracy of the mass
function and halo bias. We show the required precision in $f_i$ and
$g_i$ that can avoid $10\%$ degradation in the FoM.  The required
precision is inversely proportional to square root of the survey
area. In the context of our models, optical surveys require higher
accuracy than SZ surveys because they yield larger cluster counts and
thus have smaller {\em statistical} errors.  We find that this result
is fairly insensitive to the maximum depth of the survey and the
assumptions of the observable--mass distribution.  We mark as a star
the estimated lower limit of the required $\sigma_{\rm f}$ for a mass
proxy with 5\% scatter, which is relevant to X-ray surveys. }
\label{fig:Area}
\end{figure}

\begin{figure*}[t!]
\epsscale{1.2}
\plotone{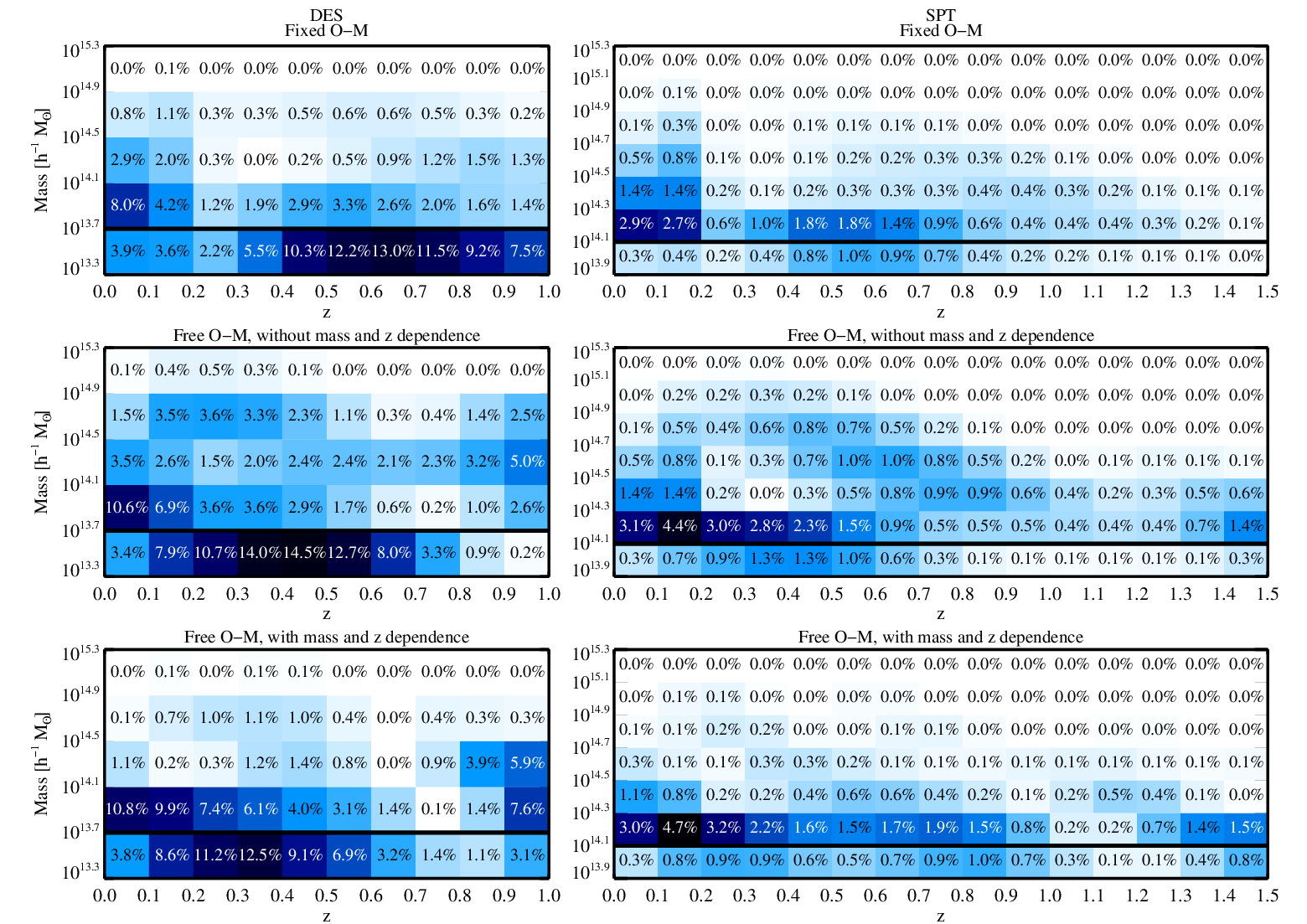}
\caption{ Relative importance of $f_i$ in different mass (vertical
axes) and redshift (horizontal axes) bins for a DES-like survey ({\em
left}) and an SPT-like survey ({\em right}).  The number in each bin
reflects the percentage improvement in the FoM that results from
tightening the prior on the mass function in that bin (from $10\%$ to
$1\%$ for DES and from $30\%$ to $3\%$ for SPT), and the shading
scales linearly with the number.  The top panels correspond to a fixed
observable--mass relation with known mean and scatter; the middle
panels correspond to a free observable--mass relation in which the
mean and scatter are self-calibrated but are not mass or redshift
dependent; and the bottom panels correspond to a free observable--mass
relation that has mass- and redshift-dependent mean and scatter
specified by the parameters of Equation~(\ref{eq:scatter}).  The
horizontal thick line shows the observable threshold; the low-mass
bins near this threshold are the most important because of the high
cluster counts in these bins. The patterns in the redshift dimension
are largely determined by a combination of the CMB prior, the clusters
counts, and the degeneracy between $f_i$ and scatter.}
\label{fig:Bins}
\end{figure*}

We specify different surveys by an observable threshold $M_{\rm th}$,
a characteristic choice of observable--mass distribution, a maximum
survey redshift $z_{\rm max}$, and a survey area $A$.  For each survey
assumption, we choose a bin size that is narrow enough to ensure no
loss of information due to binning, and we test the convergence of
parameter constraints with bin size.  We consider two broad classes of
survey that differ in the method by which clusters are identified.
First we consider typical ``optical cluster survey'' parameters with a
relatively low observable threshold and a relatively broad dispersion
in the observable--mass distribution.  Second, we consider typical
``SZ cluster survey'' parameters with a relatively high observable
threshold but a comparably small observable--mass dispersion.  The
details of these classes are as follows.
\begin{itemize}
  
  \item[1.] Optically selected clusters: $\sigma_{\rm Fid}=0.4$,
  $M_{\rm th}=10^{13.7}\ \hiMsun$, 8 bins in
  $\Mobs$ with width $\Delta\log_{10}M_{\rm obs}=0.2$, and
  $z_{\rm max}=1$.
  
  \item[2.] SZ-selected clusters:
  $\sigma_{\rm Fid}=0.2$, $M_{\rm th}=10^{14.1}\ \hiMsun$, 12
  bins in $\Mobs$ with width $\Delta\log_{10}M_{\rm obs}=0.1$, and
  $z_{\rm max}=1.5$.
  
\end{itemize}
We specifically consider a ``DES-like'' survey of optically selected
clusters over $5000 \deg^2$ of sky and an ``SPT-like'' survey of
SZ-selected clusters over $2000 \deg^2$.  We will discuss results
relevant for X-ray surveys, but we do not present distinct
calculations for the X-ray case.  As we will describe in the next
section, we find that X-ray surveys drive requirements similar to or
less stringent than those of SZ surveys.

We characterize the statistical power of galaxy cluster surveys using
the figure of merit (FoM) proposed in the Report of the Dark Energy
Task Force (DETF; \citealt{Albrecht06}):
\begin{equation}
{\rm FoM}=1/\sqrt{{\rm det}\ Cov(w_0,w_a)}=[\sigma(w_a)\sigma(w_p)]^{-1} \ ,
\end{equation}
where $w=w_0+(1-a) w_a$, and $w_p$ is calculated at the pivot redshift
at which $w$ is best constrained.  We assume a fiducial cosmology
given by the WMAP5 best-fit \citep{Komatsu09} with parameters: $w_0 =
-1,\ w_a =0,\ \Omega_{\rm DE} = 0.726,\ \Omega_{\rm k}=0,\ \Omega_{\rm
m}h^2=0.136,\ \Omega_{\rm b}h^2=0.0227,\ n_{\rm s}=0.960,\
\Delta_\zeta = 4.54\times 10^{-5} $ at $k = 0.05 {\rm Mpc^{-1}}$.  We
use a Planck prior Fisher matrix provided by Z. Ma and W. Hu (2008,
private communication).  This cosmic microwave background (CMB)
information is included in all of our calculations.

In what follows, we quote our results relative to the FoMs that may be
attained in the limit of perfect predictions for the mass and bias
functions of halos.  For each survey, the FoM in this limit can span a
range between a fixed observable--mass relation and an
observable--mass relation in which all eight nuisance parameters must
be self-calibrated.  For a DES-like survey, these baseline figures of
merit are ${\rm FoM}=17.1$ (self-calibrated observable--mass relation)
and ${\rm FoM}=139$ (fixed observable--mass relation).  In the case of
an SPT-like survey, the baseline figures of merit range from ${\rm
FoM}=5.7$ (self-calibrated observable--mass relation) to ${\rm
FoM}=42.4$ (fixed observable--mass relation).

\section{Results}
\label{sec:results}

\subsection{General Mass Function and Halo Bias Requirements}

We begin by calculating the degradation in dark energy constraints due
to uncertainty in the mass function and halo
bias. Figure~\ref{fig:f_g} shows the ratio of the degraded FoM with
respect to the fiducial FoM with perfectly known mass function and
halo bias.  We assume that all of the $f_i$ (see Equation
(\ref{eq:fi})) have the same prior, $\sigma_{\rm f}$, and all of the
$g_i$ (see Equation (\ref{eq:gi})) have the same prior, $\sigma_{\rm
g}$.  The left panels correspond to a DES-like survey, and the right
panels correspond to an SPT-like survey.

The top panels of Figure~\ref{fig:f_g} assume no prior constraints on
the observable--mass distribution. For a DES-like survey, less than
$10\%$ degradation in the FoM requires about percent-level precision
on the mass function and halo bias.  Note that the FoM is only
fractionally degraded when the mass function uncertainties increase by
an order of magnitude.  For example, when the uncertainty is as high
as $10\%$, the FoM degradation is approximately $70\%$.  On the other
hand, for an SPT-like survey, the FoM is less sensitive to mass
function and halo bias uncertainties.  This is because an SPT-like
survey has a higher observable threshold and smaller sky coverage so
that the expected cluster counts are lower and the {\em statistical}
errors are larger.

The bottom panels of Figure~\ref{fig:f_g} assume that the
observable--mass distribution is perfectly constrained.  Under this
assumption, the requirement for mass function predictions is slightly
less stringent than the case of a free observable--mass distribution
shown in the top panels.  However, the requirement for the halo bias
predictions is significantly less strict.  This behavior stems from
including the information from sample variance for self-calibrating
the observable--mass distribution.  When the observable--mass
distribution is uncertain, both the mass function and halo bias need
to be well known to avoid degrading the power of self-calibration.  On
the other hand, if the observable--mass distribution is well known,
the requirement on the halo bias is markedly less stringent because
the variance in counts is no longer needed to calibrate this relation.
Nevertheless, the small statistical errors on dark energy in the case
of a well-known observable--mass distribution mean that stringent mass
function predictions are still necessary.

We note that although the top panels and the bottom panels show
similar fractional degradations in the FoM, they have very different
FoM values and thus very different absolute degradations.  When the
observable--mass distribution is well known, uncertainties in the mass
function and halo bias will become the obstacle to achieve precision
cosmology.  On the other hand, when the observable--mass distribution
is unknown, constraining the observable--mass distribution will be
more effective to improve the FoM than constraining the mass function
and halo bias.  For detailed comparisons of uncertainties in the
observable--mass distribution and those in the mass function and halo
bias, we refer the reader to Figures 2 and 3 in \citet{Cunha09MF}.

It is worth reiterating that we have treated the mass function and
halo bias as completely independent, despite physical considerations
by which these quantities should be linked.  For example,
\citet{Manera09} have shown that the peak-background split approach
\citep{Kaiser84,MoWhite96,ShethTormen02} can be used to predict halo
biases to $\sim 5\%$ given a halo mass function.  Comparing this to
the results of Figure~\ref{fig:f_g}, an SPT-like survey can tolerate a
$\sim 5\%$ uncertainty in the halo bias when the mass function is
known, for all practical purposes, perfectly.  This indicates that it
is not necessary to consider halo bias as an independent set of
nuisance parameters in this case.  Even the requirements of DES are
not very far from this $\sim 5\%$ level of precision, suggesting that
large degradations due to an uncertain halo bias when the mass
function is well known are unlikely to happen.

In Figure~\ref{fig:Area}, we show the dependence of calibration
requirements on survey area.  To be specific, we calculate the
required $\sigma_{\rm f}$ and $\sigma_{\rm g}$ that correspond to a
$10\%$ degradation in the FoM at each value of survey area $A$.  In
cases where we compute the required $\sigma_{\rm f}$, we assume
perfect knowledge of the halo bias; we assume the converse when
computing the required $\sigma_{\rm g}$.  For the results shown in
Figure~\ref{fig:Area}, we have assumed that there are no priors on the
nuisance parameters of the observable--mass distribution.

The required $\sigma_{\rm f}$ depends upon the survey area $A$ roughly as
$\sigma_{\rm f} \propto 1/\sqrt{A}$ for the following reason.  In the
Fisher matrix, the information from the data scales as $A$, and the
information from priors scales as $\sigma_{\rm f}^{-2}$; therefore,
the scaling of $\sigma_{\rm f}$ is expected from requiring comparable
information from both data and priors.  The same reasoning applies for
$\sigma_{\rm g} $.  When comparing optical and SZ surveys, we can see
that SZ surveys require less accuracy in the mass function and halo
bias at fixed sky coverage because of the higher observable
thresholds.

We find this result to be relatively insensitive to the maximum survey
redshift once $z_{\rm max} \gtrsim 0.6$.  Above this redshift, when we
include higher-redshift bins, the increase of dark energy information
is not as rapid as in the case of low redshift.  In addition, when
redshift increases, we introduce more nuisance parameters, and these
parameters need no higher precision than those at low redshift.  We
also find that the results of Figure~\ref{fig:Area} depend only weakly
on the scatter and the assumed priors for the observable--mass
relation over a wide range of reasonable values.

We do not explicitly explore the requirement for an X-ray cluster
survey because it can be estimated from our SZ results by considering
the scatter and observable threshold.  Since X-ray clusters tend to
have smaller scatter \citep[e.g.][]{Mantz09b} than SZ clusters, we
calculate the case of a full-sky survey with a scatter $0.05$ and a
constant observable threshold $M_{\rm th}=10^{14.1}\ \hiMsun$. The
requirement on $f_i$ is about $0.5\%$, which is very close to the SZ
result.  We mark this estimate as a star in Figure~\ref{fig:Area}.  On
the other hand, X-ray clusters tend to have a higher observable
threshold than SZ clusters at high redshift.  As we have shown, higher
observable threshold requires less stringent constraints.  Therefore,
we claim that the requirement for a full-sky X-ray cluster survey will
be less stringent than $0.5\%$.


\subsection{Comparing Bins}

Next, we address the relative importance of the accuracy in the
predictions for the halo mass function at different masses and
redshifts.  There are a number of methods that could be used to
describe the relative importance of different halo masses and
redshifts.  We quantify the relative importance as follows.  We begin
with a very nonrestrictive prior on the mass function at all masses
and redshifts.  We then choose a single $f_i$ and tighten the prior on
this individual parameter.  Subsequently, we compute the percentage
improvement in the dark energy FoM with the more restrictive prior on
the mass function in the single bin.  We repeat this procedure for all
mass and redshift bins and display the relative FoM improvements in
Figure~\ref{fig:Bins}.  In each panel, the horizontal axis represents
redshift bins, and the vertical axis represents mass bins.  For a
DES-like optical survey (left panels), we start with a $10\%$ prior on
$f_i$ and improve it to $1\%$; for an SPT-like SZ survey (right
panels), we start with a $30\%$ prior and improve to $3\%$.

The relative importance of different bins does depend upon the
assumptions of the observable--mass distribution.  We show the results
for three different cases in Figure~\ref{fig:Bins}.  In the top
panels, we assume a fixed observable--mass relation.  In the middle
panels, we assume an observable--mass relation in which only $\ln M_0$
and $B_0$ of Equation~(\ref{eq:scatter}) are free to vary.  In this
case, the observable--mass distribution has a free mean and a free
scatter, but these have no mass or redshift dependence.  Lastly, in
the bottom panels, all eight parameters in Equation~(\ref{eq:scatter})
are free to vary so that the observable--mass relation may have
significant mass and redshift dependence.

One common trend for these different assumptions on the
mass--observable relation is that the low-mass bins tend to be more
important than the high-mass bins when calibrating the mass function.
Because of the steepness of the mass function, the low-mass bins
contain more statistical power than the high-mass bins.  In addition,
because of the scatter in the observable--mass distribution, accurate
predictions at the low mass end are needed to account correctly for a
potentially large number of high-$\Mobs$ clusters that up-scatter from
relatively low true masses.

In the case of the fixed observable--mass relation, the pattern in
redshift is driven by both the assumed CMB prior on cosmological
parameters and the cluster counts.  With the CMB prior,
lowest-redshift bins tend to provide the most complementary
information.  Consequently, the low-redshift bins are highlighted as
particularly important to the calibration of the theoretical
predictions. On the other hand, the low-redshift bins have small
counts due to the small volume (in our DES-like survey, the peak in
the number of clusters per redshift interval occurs near $z \approx
0.7$); therefore, some of the higher-redshift bins also have a strong
influence on the dark energy constraints.

Comparing a DES-like survey and an SPT-like survey, we find that the
lowest-redshift bins above the observable threshold (shown as a
horizontal thick line) are highlighted in both cases.  On the other
hand, in the case of DES, the mass bins below the observable threshold
are also highlighted because of the large scatter.  These bins
correspond to the greatest halo counts, and the up-scattering of halos
from these bins will be a significant fraction of the cluster sample.
Therefore, in the presence of a large scatter, the accuracy of the
mass function below the observable threshold will impact the dark
energy constraints.  We also find that this pattern depends slightly
on the binning in $\Mobs$; when the observational bin is very wide,
the importance of the mass function below the observable threshold is
less significant because the up-scattering of halos near the threshold
is not well resolved.  However, as we have already discussed, it is
most beneficial to bin finely in observations if the observable--mass
scatter can be understood even in a parameterized manner, because this
results in maximum cosmological information.

The other two rows shown in Figure~\ref{fig:Bins} allow for additional
parameter freedom in the observable--mass relation that must be
calibrated from the survey data itself.  For each survey assumption,
the structure in the two lower panels of Figure~\ref{fig:Bins} is more
complex than in the top panel.  One general difference is that the
low-redshift bins are relatively more important when the
observable--mass relation must be calibrated, and this occurs
primarily for two reasons.  First, the low-redshift bins provide
complementary information to the CMB prior.  Second, the degeneracy
between the mass function and the scatter is stronger at low redshift
because the mass function is a shallower function at low redshift.

As a result, it is most important to set stringent priors on the mass
function at low redshift and low mass.  The relative importance of
low-mass bins depends on the scatter of the observable--mass
distribution.  In the most general case of a mass- and
redshift-dependent observable--mass distribution, new parameter
degeneracies emerge but the general pattern is altered only slightly.

\section{Conclusions}
\label{sec:conclusion}

We have studied the impact of theoretical uncertainties in predictions
of the halo mass function and halo bias on dark energy constraints
from upcoming cluster surveys. Our primary conclusions are as follows.

\begin{enumerate}

\item Inaccuracy in the shape of the mass function at the $20\%$ level
  (similar to the current state of theoretical uncertainty) can lead
  to significant systematic errors in dark energy parameter
  inferences.

\item For near-term cluster surveys like DES, the mass function must
  be predicted at approximately 1\% accuracy to avoid more than 10\%
  degradation in the resulting dark energy FoM.
  Similarly, the halo bias should be predicted with a precision of
  approximately 5\%.  The current state of uncertainty in these
  functions could decrease the FoM of a DES-like survey by
  a factor of $\sim 2$.  Requirements are generally less restrictive
  for SZ and X-ray efforts.

\item A future optical survey over a significant fraction of the sky,
  like LSST, will require the most stringent constraints on the
  theoretical predictions.  In this case, the mass function and halo
  bias must be computed with an accuracy of $\sim 0.5\%$ and $\sim
  1\%$ respectively in order to guarantee that the theoretical
  uncertainty in these quantities is a negligible contributor to the
  dark energy error budget.  This represents a practical limit to the
  accuracy with which these quantities will need to be computed in
  order to interpret future survey data.

\item Precise prediction of the mass function at the low masses close
  to the observable threshold and low redshifts that will be used in
  the analysis of survey data is the most beneficial to improve dark
  energy constraints.

\end{enumerate}

We have considered the influence of theoretical uncertainties in
comparison only to statistical errors on forthcoming measurements;
however, additional systematic errors that will be present at
different levels in different types of cluster surveys will make the
demands on theoretical mass functions somewhat less restrictive.  For
example, the systematic errors due to inaccurate modeling of the
observable--mass distribution \citep[e.g. non-Gaussian
scatter,][]{Cohn07, Shaw09}, an inaccurate calibration of completeness
or false detection rate, and large errors in photometric redshift
estimates will all increase errors on dark energy parameters and
reduce the relative influence of uncertainties in predicted halo mass
functions and biases.  The requirements we advocate are relevant when
these other known sources of error do not dominate the error budget.
A robust interpretation of our calculations is that the requirements
we quote render the error due to inaccurate predictions of halo
abundances unimportant, but less stringent requirements may be
adequate when systematic errors are large.  However, we did show that
the theoretical requirement is relatively insensitive to knowledge of
the observable--mass relation when this relation is well described by
a Gaussian distribution.  This suggests that the requirements may not
be considerably loosened if systematics are moderate.

Dark matter simulations currently have $\sim 5$\% uncertainties in
predicting mass functions for fairly standard cosmological models,
even when the dark energy is {\em fixed} to a cosmological constant
\citep{Tinker08}.  To achieve sub-percent level predictions in dark
matter simulations requires a more careful consideration of
systematics in halo finding \citep{Heitmann05} and initial conditions
\citep{Crocce06}, as well as a more careful study of the
non-universality of the halo mass function
\citep{Tinker08,Robertson09}.  This will require simulations of a
wider range of cosmological models that go beyond the standard cold
dark matter with cosmological constant model.  Reasonably exhaustive
simulation programs are beginning
\citep[e.g.][]{Desjacques09,Maggiore09c,LamSheth09,Martino09,Pillepich08,Casarini09,Jennings09,
Grossi09, Alimi09}. Even more challenging still will be understanding
the systematic uncertainties induced by baryonic physics.  Recent
studies have shown that these effects can result in significant
deviations in the halo number density and that these effects are a
function of the halo mass scale \citep{Rudd08,Stanek08}.  Although
meeting the stringent constraints we advocate may be quite
challenging, pursuing further accuracy in theoretical predictions
along with controlling various systematics in clusters surveys
promises to continue to improve our knowledge of dark energy.

\acknowledgments

We are grateful to Suman Bhattacharya, Michael Busha, Bob Cobb, Carlos
Cunha, August Evrard, Salman Habib, Andrew Hearin, Katrin Heitmann,
Arthur Kosowsky, Zarija Lukic, Daisuke Nagai, Doug Rudd, Alexia
Schulz, Ravi Sheth, Jeremy Tinker, and Martin White for useful
conversations and comments on the manuscript.  We are particularly
grateful to Carlos Cunha and August Evrard for sharing their work with
us prior to publication and discussing their results with us at
considerable length.  We also thank the anonymous referee for helpful
comments.  H.W. and A.R.Z. thank the organizers of the 2009 Santa Fe
Cosmology Workshop, during which a significant amount of this work was
performed.  H.W. and R.H.W. received support from the U.S. Department
of Energy under contract number DE-AC02-76SF00515 and from Stanford
University.  A.R.Z. thanks the Michigan Center for Theoretical Physics
at the University of Michigan for supporting an extended visit during
which some of this work was conducted.  A.R.Z. is funded by the
University of Pittsburgh, by the National Science Foundation through
grant AST 0806367, and by the Department of Energy.

\bibliography{ms}
\end{document}